\documentclass[10pt, final, journal,letterpaper, oneside, twocolumn]{IEEEtran}
\usepackage{doi}
\usepackage{fancyhdr}
\usepackage{xcolor}
\usepackage{graphicx}
\ifCLASSOPTIONcompsoc
\usepackage[caption=false,font=normalsize,labelfon
t=sf,textfont=sf]{subfig}
\else
\usepackage[caption=false,font=footnotesize]{subfig}
\usepackage{dcolumn}
\usepackage{bm}
\usepackage[mathlines]{lineno}
\usepackage[utf8]{inputenc}
\usepackage[T1]{fontenc}
\usepackage[english]{babel}
\usepackage{cite}
\usepackage{amsmath}
\graphicspath{./}

\begin{document}

\title{Symmetric Traveling Wave Parametric Amplifier}
\author{\IEEEauthorblockN{Alessandro Miano\IEEEauthorrefmark{1} \IEEEauthorrefmark{3} and
		Oleg A. Mukhanov\IEEEauthorrefmark{2}  \IEEEauthorrefmark{3}}\\
	\IEEEauthorblockA{\IEEEauthorrefmark{1}Dept. of Physics "Ettore Pancini", University of Naples "Federico II", via Cinthia 21, 80126 Naples - Italy}\\
	\IEEEauthorblockA{\IEEEauthorrefmark{2} HYPRES-SeeQC, Inc., 175 Clearbrook Road, Elmsford, NY 10523 - USA}\\
	\IEEEauthorblockA{\IEEEauthorrefmark{3} SeeQC-eu, Via dei Due Macelli 66, 00187 Rome - Italy}}
\maketitle
\thispagestyle{fancy}
\begin{abstract}
	We developed and experimentally tested a Symmetric Traveling Wave Parametric Amplifier (STWPA) based on Three-Wave Mixing, using the new concept of a Symmetric rf-SQUID. This allowed us to fully control the second and third order nonlinearities of the STWPA by applying external currents. In this way, the optimal bias point can be reached, taking into account both phase mismatch and pump depletion minimization. The structure was tested at 4.2K, showing a 4GHz bandwidth and a maximum estimated gain of 17dB.
\end{abstract}
\section{Introduction}
\IEEEPARstart{S}{calable} superconducting quantum computing demands non dissipative, quantum limited and wide bandwidth cryogenic amplifiers. In order to improve scalability, wide bandwidth amplifiers with reasonable gain allow an efficient implementation of readout multiplexing when multiple qubits can be sensed with a single amplifier. 

Josephson Parametric Amplifiers (JPAs) were introduced to meet these needs, and proposed in many different designs \cite{jpc,flux_jpa,beyond_gb,mutus}. While achieving great gain and noise performance, JPAs cannot guarantee a wide instantaneous bandwidth for many-qubit multiplexing, because of a lumped-circuit based design.
Good gain-bandwidth results are achieved by exploiting nonlinear kinetic inductance of NbTiN \cite{vissers}, but lengths in the order of meters are required, leading to difficult integration.
A recent approach is to build arrays of small Josephson based elementary cells, forming a so-called \emph{Josephson Metamaterial}. Josephson Metamaterials show an overall tunability comparable to its individual components \cite{jos_meta}, allowing external control of main parameters \cite{castellanos}.
Based on this approach, the Josephson Traveling Wave Parametric Amplifier (TWPA) was introduced \cite{yaakobi} as a nonlinear transmission line containing Josephson junctions. This novel device is achieving \cite{fract_twpa,bell,zorinexp,white,macklin} many GHz bandwidth, high gain and near quantum-limited performances.

As a parametric amplifier, TWPA transfers power from a strong microwave tone (pump) to a weak one (signal). With a choice of the Josephson nonlinearity responsible for parametric process, two different approaches can be used, as depicted in Fig. \ref{fig_wm}: four-wave mixing (4WM) \cite{white,macklin}, based on the third-order nonlinearity, or three-wave mixing (3WM) \cite{zorinth,zorinexp}, based on the second-order nonlinearity. 
While 4WM TWPAs achieve high gain, bandwidth and noise performance, they require {resonant circuits to work properly} \cite{white,macklin}, leading to a stopband in the gain curve. 
The recent 3WM approach, instead, is promising to achieve comparable experimental results with a smaller structure \cite{zorinexp} and external control via DC flux bias. One of the main 3WM advantages is that parametric process efficiency and phase mismatch can be independently tuned\cite{zorinth}, respectively depending on second and third order Josephson nonlinearities. { However}, with a single DC flux bias, second and third order nonlinearities cannot be independently set \cite{zorinth}. This could lead not only to challenging device design, but limits a full exploration of 3WM TWPA possibilities.

To allow independent control of both second and third order Josephson nonlinearities, we developed a three-wave mixing Symmetric Traveling Wave Parametric Amplifier (STWPA) based on a new symmetric rf-SQUID scheme. Experimental measurements of the first wafer demonstrated the correct TWPA operation at 4.2 K, showing a gain up to 17dB and a 4GHz 3dB bandwidth. In this paper, we present the basics of the TWPA with three-wave mixing and explain our symmetric approach to TWPA design showing its advantages. Then, we present the experimental setup and devices details, with experimental data from measurements. Finally, we consider possible applications of the STWPA.

\section{Three Wave Mixing TWPA}
\subsection{Previous art}
\begin{figure}
	\centering
	\subfloat[Four-wave mixing TWPA]{\includegraphics[width=0.5\columnwidth]{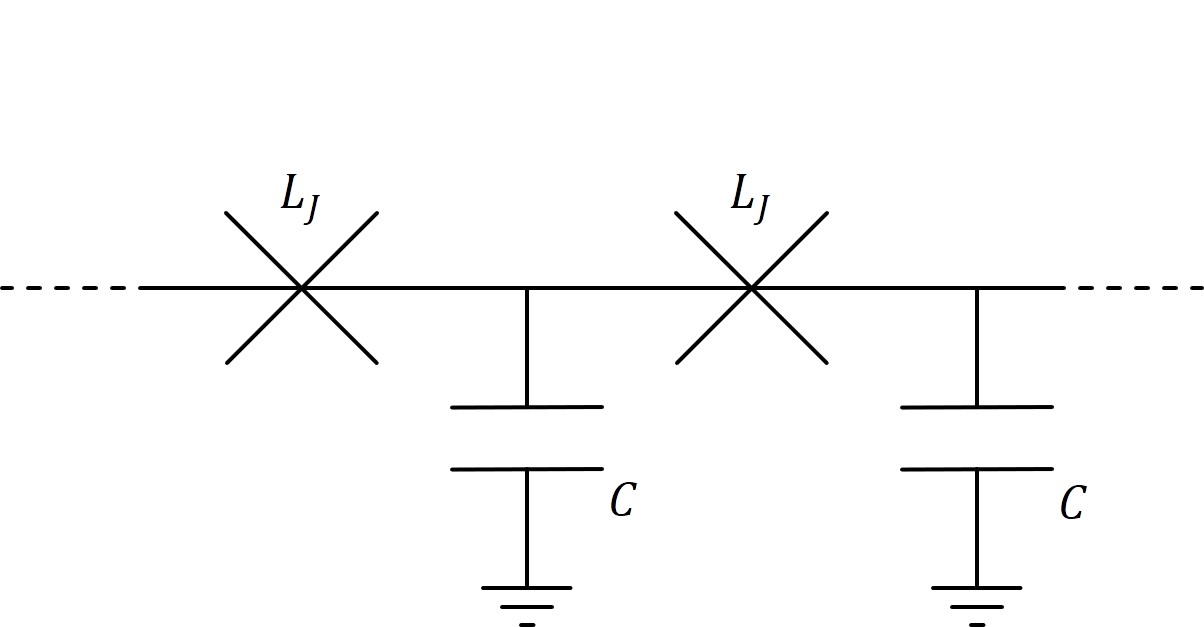}
		\label{fig_4wm}}
	\subfloat[Three-wave mixing TWPA]{\includegraphics[width=0.5\columnwidth]{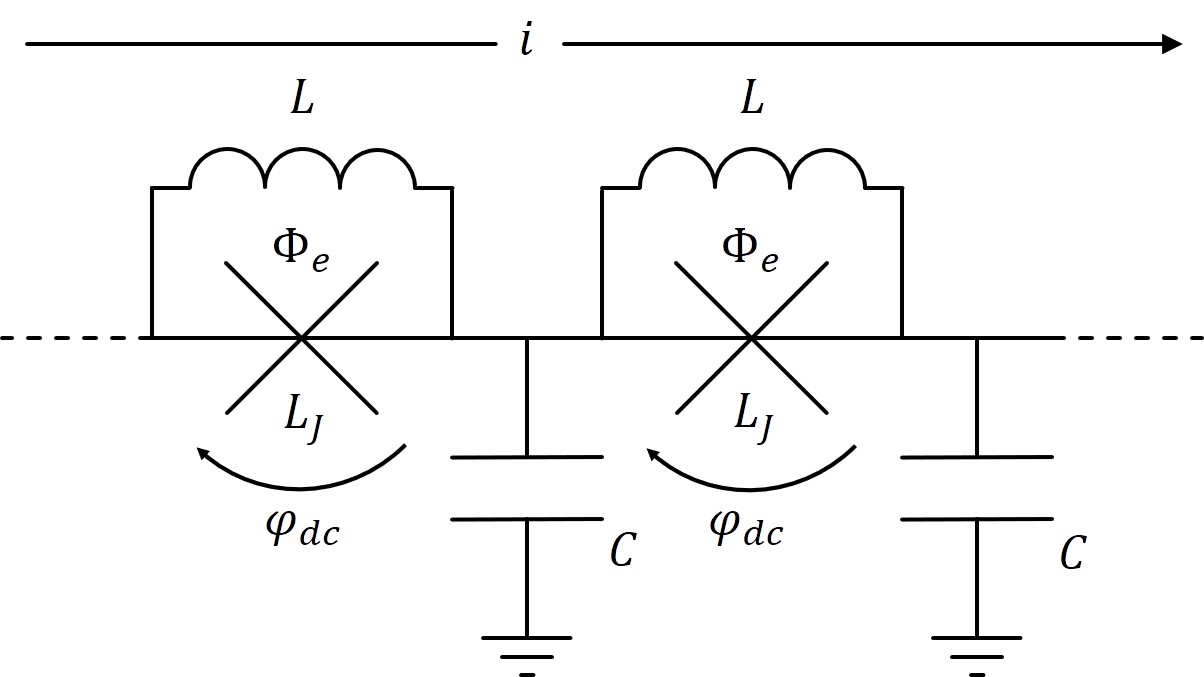}
		\label{fig_3wm}}
	\caption{Basic TWPA structures. (a) 4WM TWPA can be achieved with a Josephson junction as the inductive element $L_J$ of a lossless waveguide. (b) 3WM, instead, uses an rf-SQUID as the inductive element. A DC current $i$ gives flux bias $\Phi_e$ to each rf-SQUID, resulting in a phase bias $\varphi_{dc}$.}
	\label{fig_wm}
\end{figure}

Depending on the involved nonlinearity, the inductive part of the TWPA elementary cell can be made with different schematics (see Fig. \ref{fig_wm}).

Three-wave mixing TWPAs use { the} second-order nonlinearity in the {Josephson current relation} $j(\varphi)=I_c\sin{\varphi}$ to produce a second order stimulated parametric process \cite{zorinth,zorinexp}.
Then, three AC tones are involved: pump ($f_p$), signal ($f_s$) and idler ($f_i$). The idler tone is produced to satisfy energy conservation relation $f_p=f_s+f_i$ (for $f_p>f_s$).

To achieve a three-wave mixing process, tuning of $j(\varphi)$ around $\varphi\approx\pi/2$ is needed. This is easily obtained by application of a DC flux bias to the rf-SQUIDs, given that the \emph{dispersive condition} $\tilde{\beta}=2\pi LI_c/\Phi_0<1$ \cite{barone_squid} is satisfied, where $I_c$ is the critical current of a Josephson junction, $L$ is the shunting geometrical inductance closing the loop and $\Phi_0$ is the magnetic flux quantum.
In fact, if $\tilde{\beta}<1$, the phase-flux relation
\begin{equation}
\label{phase_flux}
\varphi+\tilde{\beta}\sin{\varphi}=2\pi\frac{\Phi_e}{\Phi_0}
\end{equation}
becomes a one-to-one correspondence, where $\Phi_e$ is the external flux concatenated to the rf-SQUID loop (see Fig. \ref{fig_3wm}).

Given the dc phase bias $\varphi_{dc}$ solution of equation \ref{phase_flux}, under the hypothesis that all the ac tones involved give small perturbations $\delta\varphi$ to the junction's phase $\varphi$, the Josephson current $j(\varphi)$ can be expressed as
\begin{align}
\begin{split}
\label{current_phase}
j(\varphi_{dc}+\delta\varphi)\simeq&I_c\sin{\varphi_{dc}}\bigg(1-\frac{{\delta\varphi}^2}{2}\bigg)\\
{+}&I_c\cos{\varphi_{dc}}\bigg(\delta\varphi-\frac{{\delta\varphi}^3}{6}\bigg).
\end{split}
\end{align}
From relation \ref{current_phase}, it's easy to see that $\sin{\varphi_{dc}}$ controls the even terms, while $\cos{\varphi_{dc}}$ controls the odd ones.

Three-wave mixing TWPA wave equation has quadratic and cubic nonlinearities, respectively, controlled by coefficients $\beta=\tilde{\beta}\sin{\varphi_{dc}}/2$ and $\gamma=\tilde{\beta}\cos{\varphi_{dc}}/6$ \cite{zorinth}.
While the second order effect is maximum when $\varphi_{dc}=\pi/2$, the same does not apply to signal amplification. 
{ In fact, solving 3WM TWPA wave equation \cite{zorinth} leads to the following expression for maximum gain
\begin{equation}
G_{max}=1+\sinh^2{\left(|\beta|A_{p0}\sqrt{4\omega_s\omega_i}\omega_p/ 4\omega_0^2\right)}
\label{max_gain}
\end{equation}
where $A_{p0}$ and $\omega_0=1/\sqrt{LC}$ are, respectively, pump amplitude at input port and waveguide cutoff frequency.

The last expression can be obtained only if, the total phase mismatch
\begin{equation}
\psi=3\frac{\omega_p^3}{8\omega_0^3}\left[\frac{4\omega_s\omega_i}{\omega_p^2}\frac{\omega_0^2}{\omega_J^2}-\gamma A_{p0}^2\right], 
\label{phase_mism}
\end{equation}

is set to zero, where $\omega_J$ is the plasma frequency of Josephson junctions.
From last equation is clear that phase mismatch cannot be set to zero if $\varphi_{dc}=\pi/2$ (i.e. maximizing $\beta$ and consequently the gain), because that would lead to $\gamma=0$. It's in principle possible to use $A_{p0}$ as a second parameter, but it controls also a negative effect like pump depletion \cite{zorinth}.
}
To solve this problem, we introduced a new concept of rf-SQUID based on a symmetric design.
\subsection{New Approach: Symmetric rf-SQUID}
\begin{figure}[h]
	\centering
	\subfloat[Symmetric rf-SQUID]{\includegraphics[height=0.8\columnwidth]{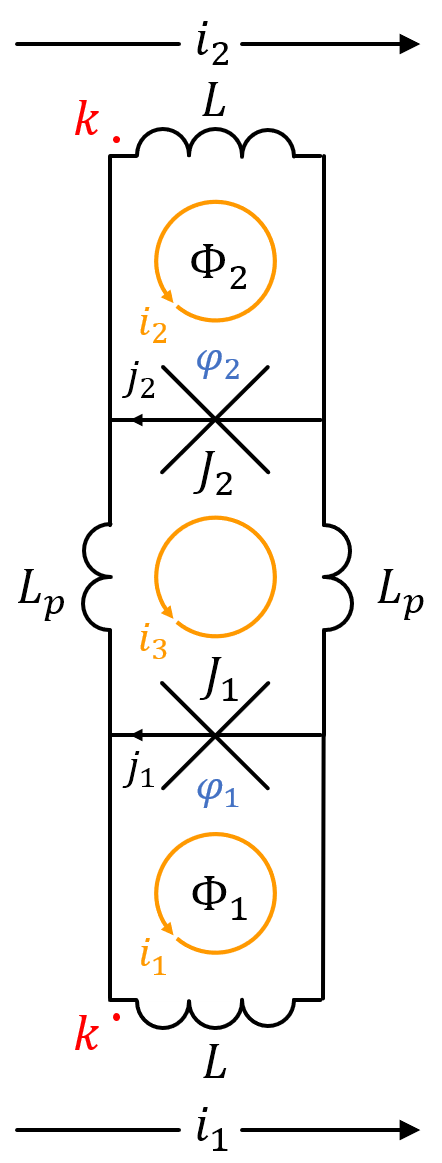}
		\label{fig:sym_rfsquid}}
	\subfloat[Symmetric TWPA]{\includegraphics[height=0.8\columnwidth]{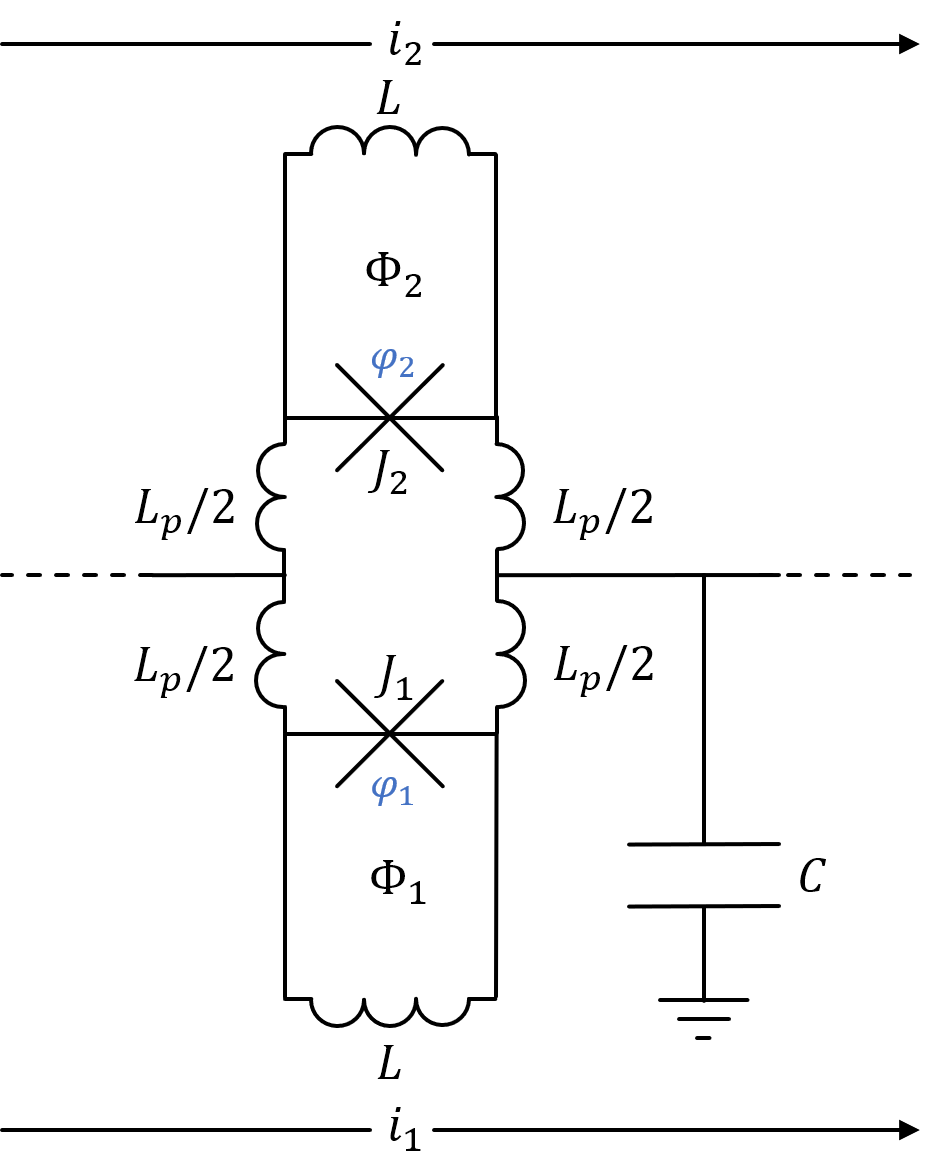}
		\label{fig:stwpa}}
	\caption{Symmetric rf-SQUID and its TWPA implementation. 
		(a) Symmetric rf-SQUID structure, made by shunting two identical rf-SQUIDs. Shunting conductors offer a parasitic inductance $L_p$. When flux bias is applied by means of DC currents $i_1$ and $i_2$, the junctions $J_1$ and $J_2$ are phase biased, respectively, with dc phases $\varphi_{1}$ and $\varphi_{2}$. (b) Symmetric TWPA (STWPA) structure. The rf line is connected at the center of the wiring inductance $L_p$, that gets split into two equal contributes $L_p/2$ for preserving design symmetry.}
\end{figure}

\begin{figure}[h]
	\centering
\subfloat[Bias phases plane]{\includegraphics[width=0.5\columnwidth]{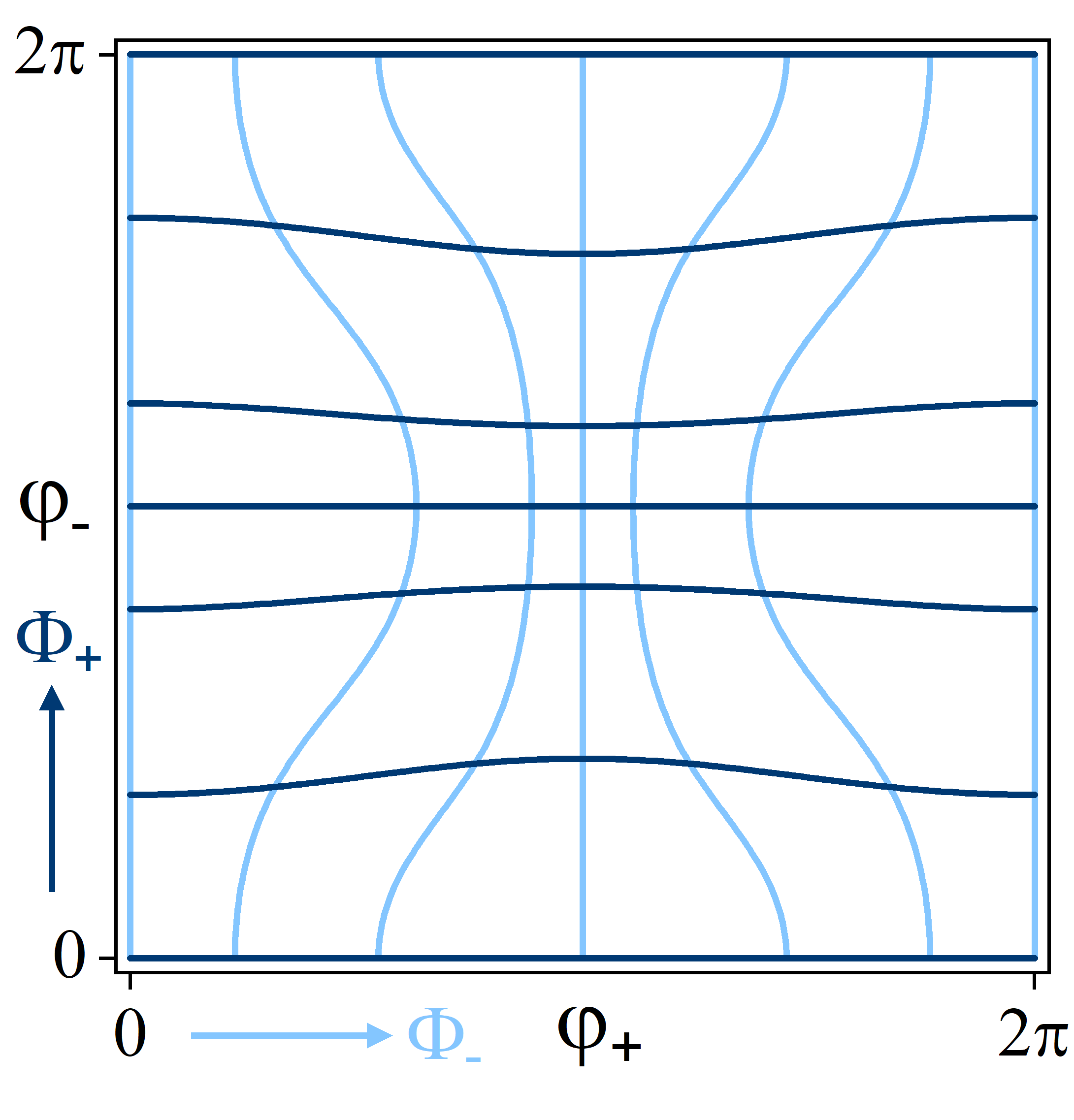}
	\label{fig:phase_plane}}
\subfloat[Coefficients plane]{\includegraphics[width=0.5\columnwidth]{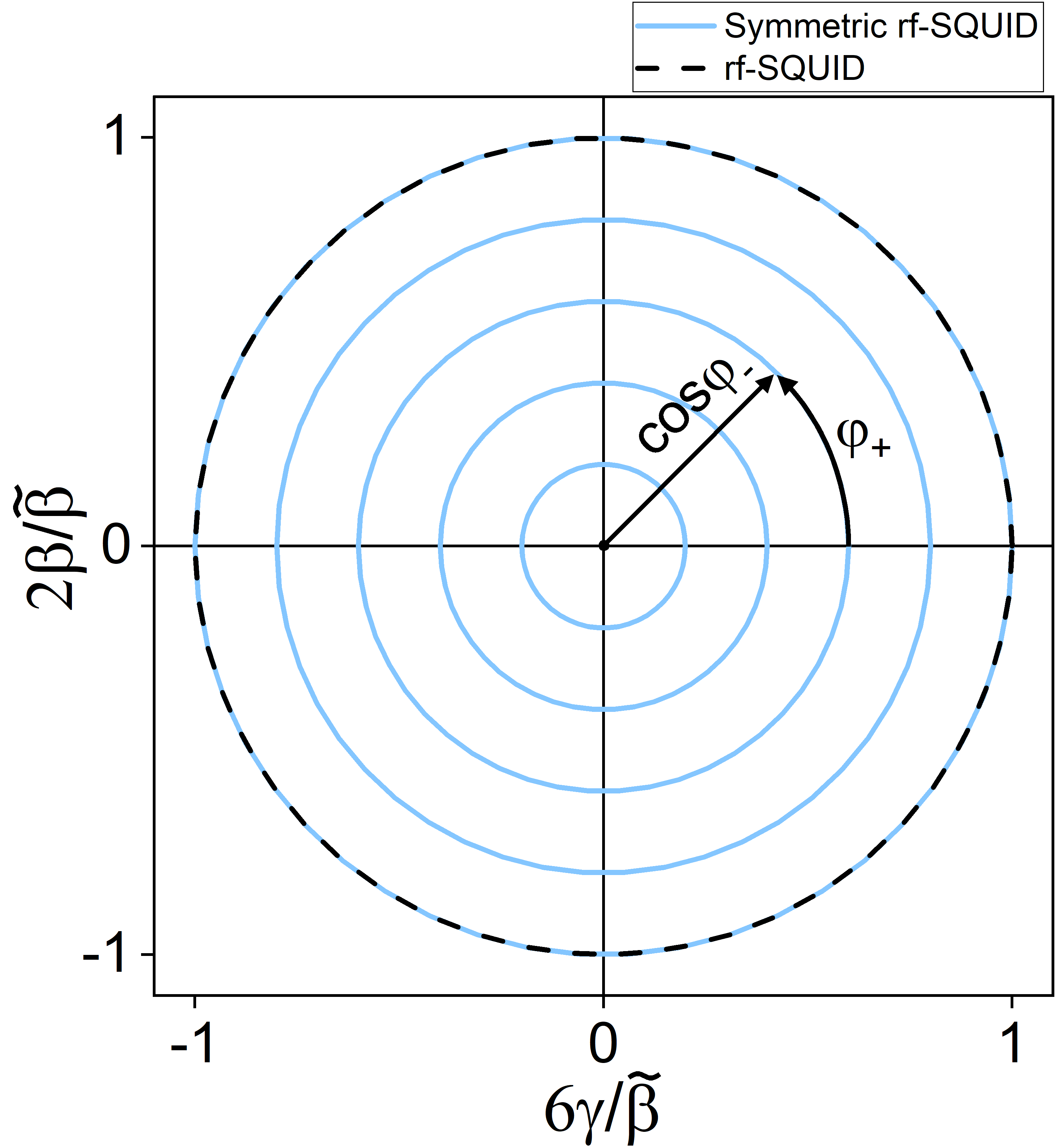}
	\label{fig:par_plane}}
	\caption{Numerical analysis of symmetric rf-SQUID biasing on $(\varphi_1,\varphi_2)$ plane and $(\beta,\gamma)$ plane. (a) Simulated constant-flux curves on the bias phases plane for $\beta=0.8$, $\alpha=5$ and $k=0$. Each { dark and light} curve represent, respectively, a phase curve for a given value of $\Phi_+$ or $\Phi_-$. (b) Normalized coefficients plane. Each {solid} circle represents, for a STWPA, the available values of $\beta$ and $\gamma$ for a given value of $\varphi_-$, when $\varphi_+\in[0,2\pi]$. The black dashed line, instead, represents the only available curve for the standard TWPA.}
\end{figure}

A symmetric rf-SQUID is obtained by shunting two identical rf-SQUIDs. Taking into account a parasitic inductance $L_p$ connecting the two rf-SQUIDs, we obtain the schematic in Fig. \ref{fig:sym_rfsquid}.
Using net currents ($i_1,i_2,i_3$ as in Fig. \ref{fig:sym_rfsquid}) method, and noticing that side nets are rf-SQUIDs and central net is a dc-SQUID, it's easy to obtain the following equations for the symmetric rf-SQUID with dc magnetic flux bias:
\begin{equation}
\label{bias_eq}
\left\{
\begin{array}{ll}
\varphi_++\tilde{\beta}(1+k)\sin{\varphi_+}\cos{\varphi_-}=2\pi\dfrac{\Phi_-}{\Phi_0}\\[10pt]
\varphi_-+\tilde{\beta}\dfrac{1-k}{1+\alpha}\sin{\varphi_-}\cos{\varphi_+}=\dfrac{2\pi}{1+\alpha}\dfrac{\Phi_+}{\Phi_0}
\end{array}
\right.
\end{equation}
where 
\begin{align}
\nonumber
\varphi_+&=\dfrac{\varphi_1+\varphi_2}{2} & \varphi_-&=\dfrac{\varphi_2-\varphi_1}{2}\\[10pt]
\nonumber
\Phi_+&=\dfrac{\Phi_1+\Phi_2}{2} & \Phi_-&=\dfrac{\Phi_2-\Phi_1}{2}\\[10pt]
\nonumber
\alpha&=\dfrac{L(1-k)}{L_p}>0 & \tilde{\beta}&=\dfrac{2\pi LI_c}{\Phi_0}
\end{align}
being $I_c$ the critical current of both junctions $J_1$ and $J_2$, supposed identical and $k$ the coupling coefficient between the two geometrical inductances of each rf-SQUID. Here, we set to zero the external flux $\Phi_3$ concatenated with the central net.
It can be shown that a sufficient condition for dispersive operation is still $\beta<1$, if $k<<1$.
Then, in dispersive mode, is possible to tune external fluxes $\Phi_1$ and $\Phi_2$ to achieve all the values for the bias phase vector $(\varphi_1,\varphi_2)\in[0,2\pi]\times[0,2\pi]$, as is clear in Fig. \ref{fig:phase_plane}.
This result allows { us} to independently tune the phases of the two junctions, adding a second bias phase degree of freedom to the rf-SQUID.
\subsection{Symmetric TWPA}
To integrate a TWPA using symmetric rf-SQUIDs, it is necessary to "cut" the shunting inductance $L_p$ at a given section, so to cascade connect the devices. To preserve the symmetry of the structure, this should be done by halving $L_p$ into two identical parts, resulting in a Symmetric TWPA (STWPA) structure as in Fig. \ref{fig:stwpa}.
Currents flowing into the two Josephson junctions of each symmetric rf-SQUID can be expressed, for small phases perturbations $\delta\varphi_1$ and $\delta\varphi_2$, as
\begin{align}
\begin{split}
\label{current_phase1}
j_1(\varphi_{1}+\delta\varphi_{1})\simeq&I_c\sin{\varphi_{1}}\bigg(1-\frac{{\delta\varphi_{1}}^2}{2}\bigg)\\
+&I_c\cos{\varphi_{1}}\bigg(\delta\varphi_{1}-\frac{{\delta\varphi_{1}}^3}{6}\bigg)
\end{split}\\
\begin{split}
\label{current_phase2}
j_2(\varphi_{2}+\delta\varphi_{2})\simeq&I_c\sin{\varphi_{2}}\bigg(1-\frac{{\delta\varphi_{2}}^2}{2}\bigg)\\
+&I_c\cos{\varphi_{{2}}}\bigg(\delta\varphi_{2}-\frac{{\delta\varphi_{2}}^3}{6}\bigg).
\end{split}
\end{align}
A relation between $\delta\varphi_1$ and $\delta\varphi_2$ can be found with a small signal analysis of the elementary cell of STWPA, resulting in the following (valid for each STWPA section)
\begin{equation}
\label{phase:variation:relation}
\frac{\delta\varphi_1}{\delta\varphi_2}=\frac{1+\alpha+\tilde{\beta}\cos{\varphi_2}}{1+\alpha+\tilde{\beta}\cos{\varphi_1}},
\end{equation}
where is clear that the two phase perturbations are proportional by means of a real constant (no phase shift) and their ratio does not depend on signal frequency.
 Then, for reasonably big values of $\alpha$, and given that the phase working point of each junction is near $\pi/2\pm\pi$, it easy to show that STWPA wave equation can be expressed in an explicit form, formally identical to the standard TWPA's one \cite{zorinth}. We want to specify that this hypothesis is only necessary to obtain a simple form for STWPA wave equation because of the $L_p$ inductance, but is not mandatory for its correct operation.
 Nonlinear terms are then given by the sum of the two nonlinear currents in relations \ref{current_phase1} and \ref{current_phase2}, producing the quadratic and cubic coefficients, respectively, given by
 
 \begin{equation}
 \label{plane_eqn}
 \begin{split}
 \beta=\frac{\tilde{\beta}}{2}\cos{\varphi_-}\sin{\varphi_+}\\
 \gamma=\frac{\tilde{\beta}}{6}\cos{\varphi_-}\cos{\varphi_+}
 \end{split}
 \end{equation}
 
 where trigonometric sum to product formulas have been used.  By taking into account the full relation \ref{phase:variation:relation}, some correction factors should be applied to relations \ref{plane_eqn} to have the most accurate description of STWPA bias properties. Anyway, relations \ref{plane_eqn} are accurate enough for description of STWPA operations and understanding experimental data.
 
 By varying $\varphi_+$ and $\varphi_-$ by means of flux bias as in equations \ref{bias_eq}, it's possible to independently set all the values for $\beta$ and $\gamma$, once $\tilde{\beta}$ has been fixed. As clear in \ref{plane_eqn} and depicted in Fig. \ref{fig:par_plane}, normalized coefficients $\beta/2\tilde{\beta}$ and $\gamma/6\tilde{\beta}$ describe circles when bias phases are varied. Each point of a circle is represented by a radius $r=|\cos{\varphi_-}|$ and and angle $\theta=\varphi_+$ ($\cos{\varphi_-}\geq0$) or $\theta=\pi+\varphi_+$ ($\cos{\varphi_-}<0$).

 \section{Methods and Experimental Results}
 \subsection{Chip Design and Measurements Setup}

\begin{figure}[h]
 	\centering
 	\includegraphics[width=\columnwidth]{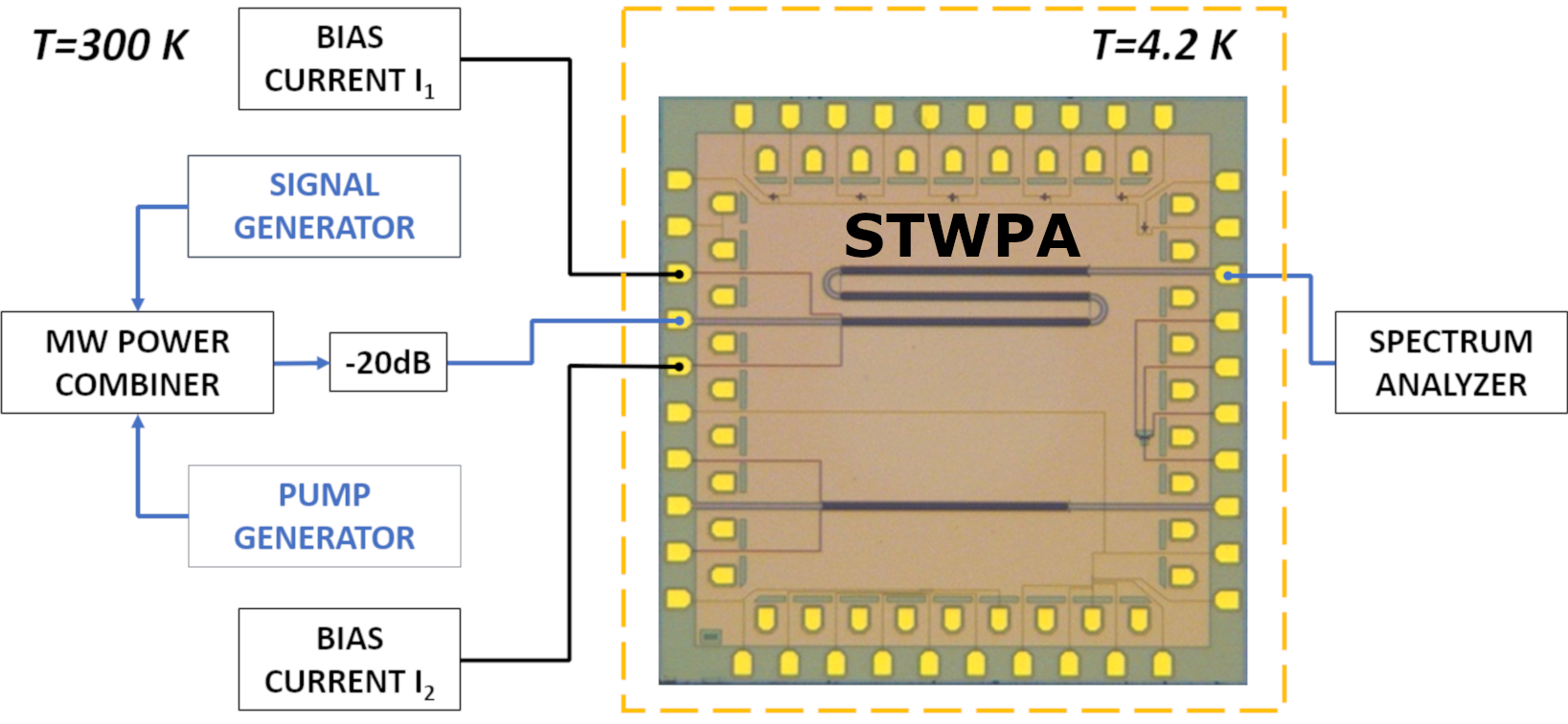}
 	\caption{Experimental setup for gain measurements with a micrograph of a $5\times5 mm^2$ chip fabricated using $100 A/cm^2$. Pump and signal tones were generated by two independent microwave generators and mixed with a power combiner. Two independent current generators were used to apply an external magnetic flux to the STWPA.}
 	\label{fig:setup}
 \end{figure}

 The symmetric rf-SQUID was simulated and optimized using \emph{AWR Microwave Office} and \emph{InductEx} softwares. Then, an optimized circuit layout was designed for HYPRES' Nb process with $J_c=100A/m^2$.
 Various test chips were designed, both for a 4.2K operation and for a mK operation (test in progress). On each chip, there are a STWPA with reference design similar to ones described in \cite{zorinth} and various diagnostic circuits, as in Fig. \ref{fig:setup}.
 
The symmetric rf-SQUID structure was made with Josephson junctions with different nominal critical current $I_c=$\{4uA, 6uA, 12uA\}. Geometrical inductances were chosen to satisfy dispersive condition for each structure.
Grounding capacitors were designed to match a $50\Omega$ impedance. Elementary cells were then made with a grounding capacitor every 6 symmetric rf-SQUID, as suggested from \cite{zorinexp}, as well as 2 and 1 each grounding capacitor. Then, the number of cells N was set around the optimal value \cite{zorinth} for each STWPA. Effective signal and pump power at amplifier input port were $\approx$ -120dBm and $\approx$ -65dB respectively. Pump frequency was around $f_p=$10.2GHz and $f_s\in$ [3.4,9.4] GHz. Bias current values were in the range of the mA.

\subsection{Experimental Verification}

To validate our structure on the basis of model and simulations, we measured signal output power as a function of the two bias currents. Bias fluxes $\Phi_+$ and $\Phi_-$ are related to bias currents by relations
\begin{equation}
\label{curr_flux_2}
\begin{split}
\Phi_+=(M_1-M_2)I_-\\
\Phi_-=(M_1+M_2)I_+
\end{split}
\end{equation}
where $M_1$ is the mutual inductances between a flux line and its correspondent symmetric rf-SQUID side, $M_2$ is the mutual inductance between a flux line and the opposite symmetric rf-SQUID side, $I_+=(I_1+I_2)/2$ and $I_-=(I_1-I_2)/2$. Combining equations \ref{curr_flux_2} and \ref{bias_eq}, we observe a strong dependence of $\varphi_+$ and $\varphi_-$, respectively, by $I_+$ and $I_-$. Then, observing from equations \ref{plane_eqn} that the bias point $(\beta,\gamma)$ has to be even (and also periodic) in $\varphi_-$ and periodic in $\varphi_+$, the same should approximately apply to $I_-$ and $I_+$.
 \begin{figure}[h]
	\centering
	\includegraphics[width=\columnwidth]{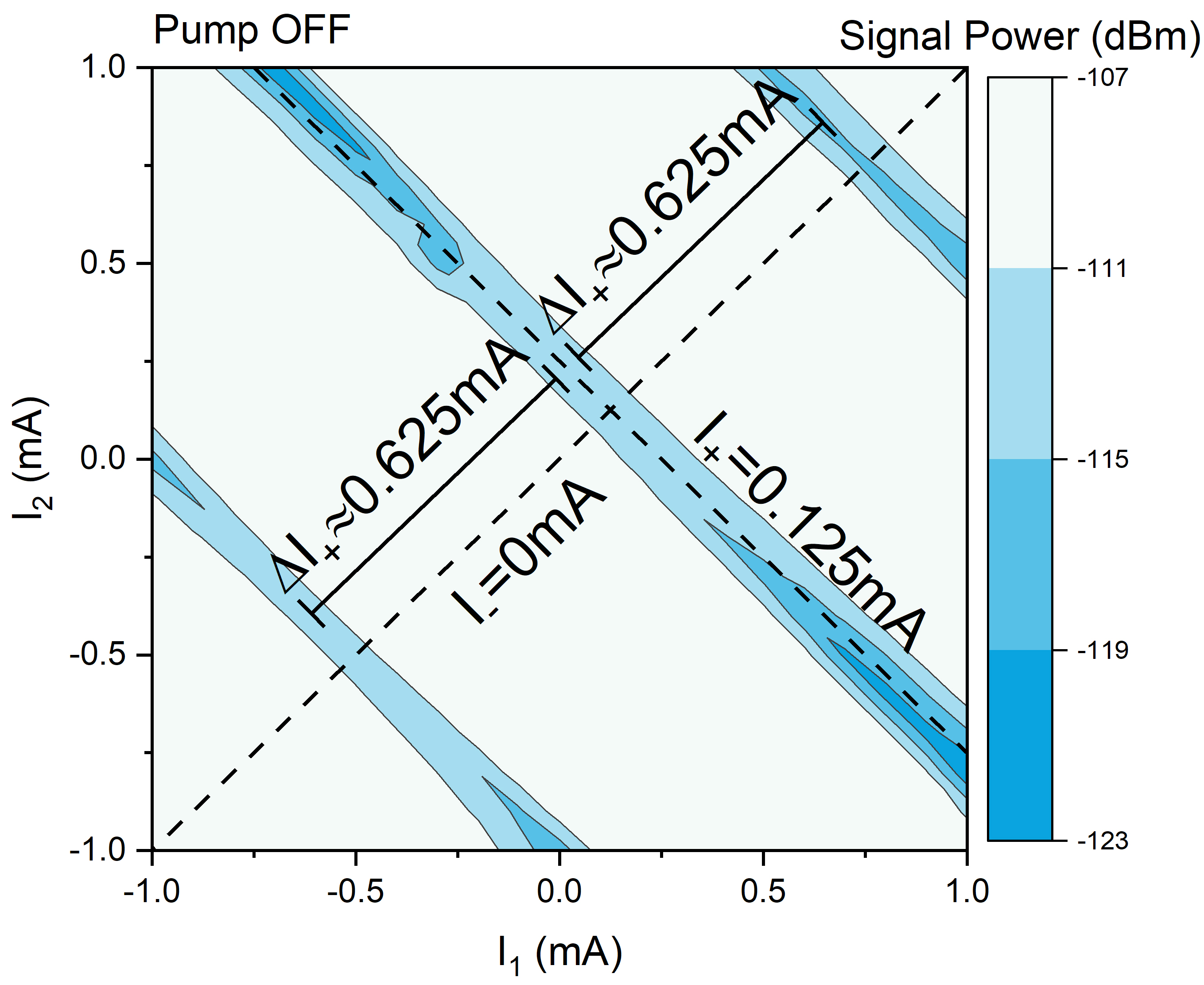}
	\caption{Signal Power vs. $I_1$ and $I_2$ with pump off. Expected behavior of symmetry within $I_-$ and periodicity within $I_+$ is clear from the image. Each diagonal segment is relative to a given value of $I_+$.}
	\label{sweep_1}
\end{figure}
This is confirmed by data shown in Fig. \ref{sweep_1}, where signal power (with pump OFF) is measured by sweeping both the bias currents. In fact, the plot is specular with respect to the $I_-=0mA$ segment, meaning that the device operation is even in $I_-$. Periodicity in $I_+$ is verified noticing that diagonal segments are equally spaced by a value of $\Delta I_+\approx0.625mA$. To locate these segments onto the coefficient plane in Fig. \ref{fig:par_plane}, we can note that the first segment appears around $I_+=0.125mA$. Then, because a full turn on the coefficient plane corresponds to $\approx 0.625mA$, the first segment corresponds, approximately, to a phase value $\varphi_+\approx2\pi/5=\pi/2-\pi/10$. This means that when bias currents are given into these segments, $\varphi_+\approx\pi/2+2n\pi$, being $n$ an integer number. So, these segments correspond to a very high second order effect, being $\beta$ maximized when $\varphi_+=\pi/2$.
\subsection{Gain Measurements}
\begin{figure}[h]
	\centering
	\includegraphics[width=\columnwidth]{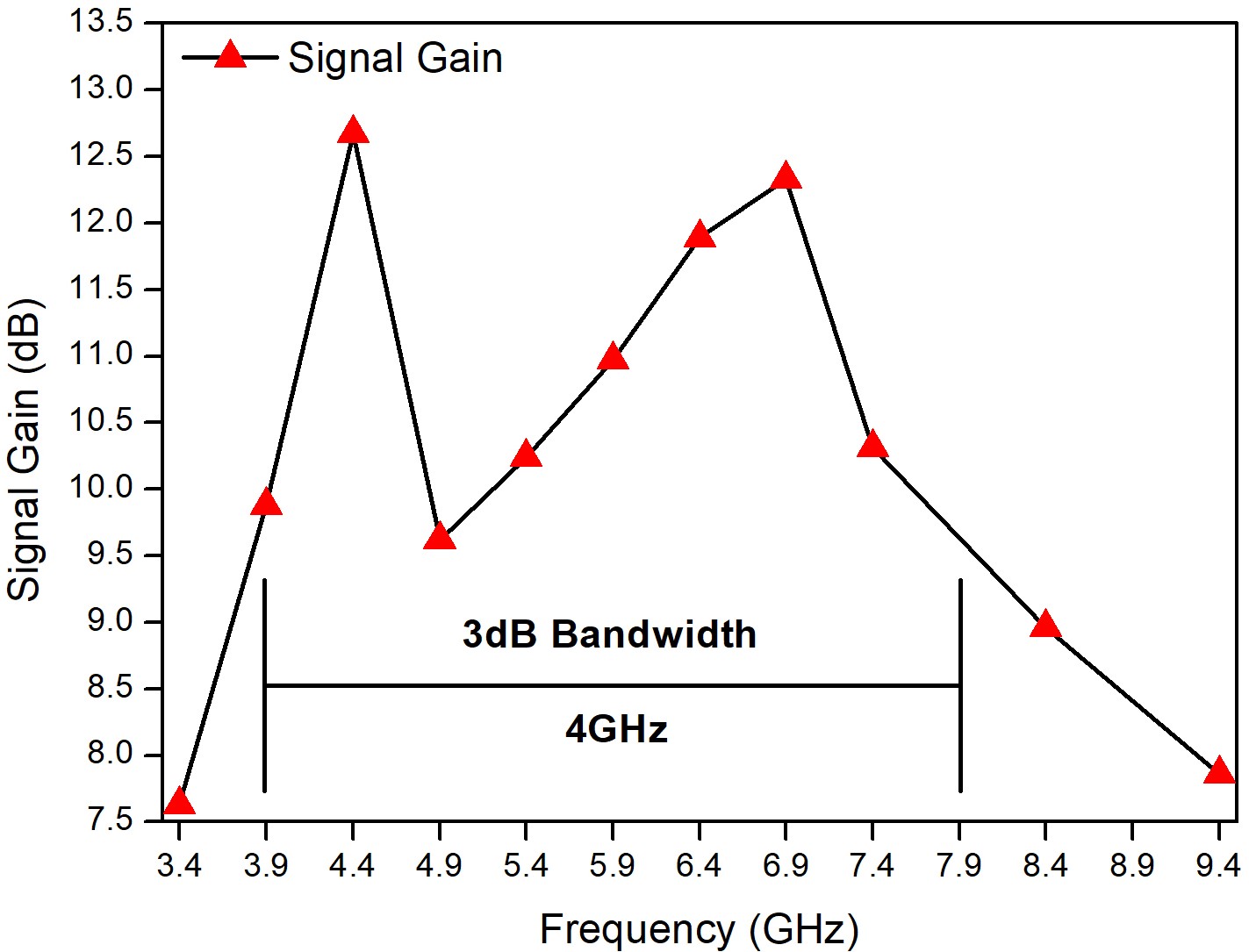}
	\caption{Measured signal gain curve as a function of signal frequency. It showed a 4GHz 3dB bandwidth with a peak value around 12.5 dB. Displayed data was measured with the STWPA biased with a same set of the experimentally optimized parameters. { The gain drop around 4.9GHz can be attributed to an unwanted mismatch due to fabrication fidelity and/or test setup mismatch affecting the gain measurement.}}
	\label{data}
\end{figure}

We measured signal gain as $G_s=P_s^{on}-P_s^{off}$, where $P_s^{off}$ and $P_s^{on}$ were, respectively, signal power (in dBm units) at output port (connected to spectrum analyzer) with amplifier off (bias currents and pump off) and with amplifier on (bias currents and pump on). Off state values could have been affected by some frequency depending mismatch, because STWPA was designed to match an impedance of $\approx50\Omega$ when flux biased around $\varphi_{1,2}\approx\pi/2$, where total rf-SQUID inductance has only the geometrical contribution. Instead, for a general phase bias, Symmetric rf-SQUID total inductance is given by
\begin{equation}
\label{rf_inductance}
L_{tot}=\frac{L}{2(1+\tilde{\beta}\cos{\varphi_+}\cos{\varphi_-})}
\end{equation} 
being $\beta\approx1$, with no bias currents applied (i.e. $\varphi_+=\varphi_-=0$), inductance per unit length is halved, resulting in an effective impedance around $Z_0/\sqrt{2}\approx 35\Omega$. Maximum deviation of OFF state signal power due to mismatch was then estimated, resulting in a maximum reflection loss of about 2dB.
By swapping bias currents and measuring the same gain, we proved the validity of our symmetric design. Gain was measured and some of the collected data is shown in figure \ref{data}, resulting in a gain >9.5 dB in a 4GHz bandwidth centered at 5.9 GHz, with a nominal peak value of 12.5dB. Moreover, by moving a little the pump power around the value used for data in figure \ref{data}, we achieved higher gain for some signal frequencies, as a 17dB gain at 7.9GHz, 11dB at 8.4GHz and 9.4GHz. At these frequency values, we estimated an error caused by mismatch of about 1dB.
\subsection{Pump Driven rf-Switch and Up-Down Converting Mixer}
\begin{figure}[h]
 	\centering
 	\subfloat[Pump Off Sweep]{\includegraphics[width=\columnwidth]{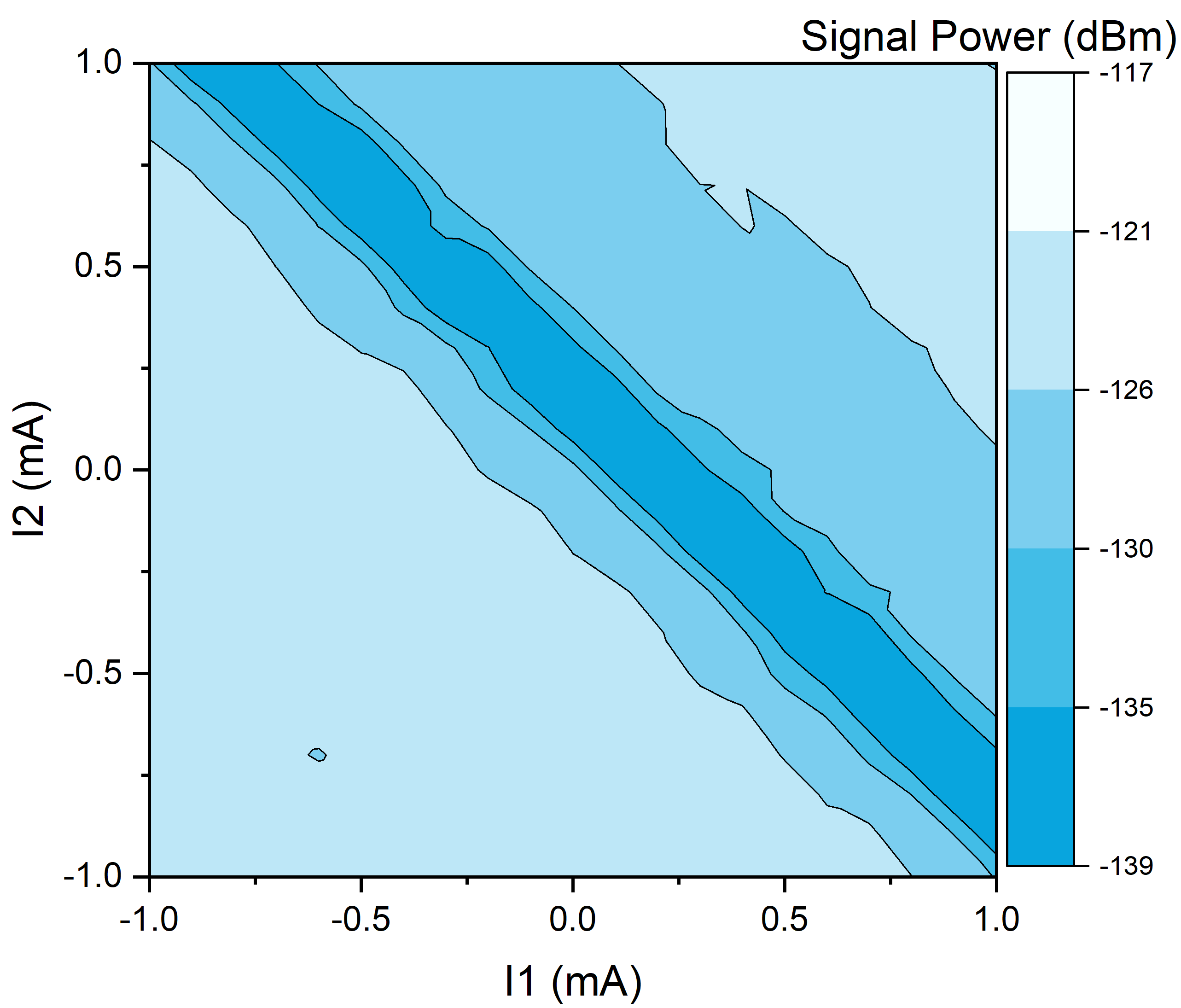}
 		\label{sweep_2_off}}
 	\hfil
 	\subfloat[Pump On Sweep]{\includegraphics[width=\columnwidth]{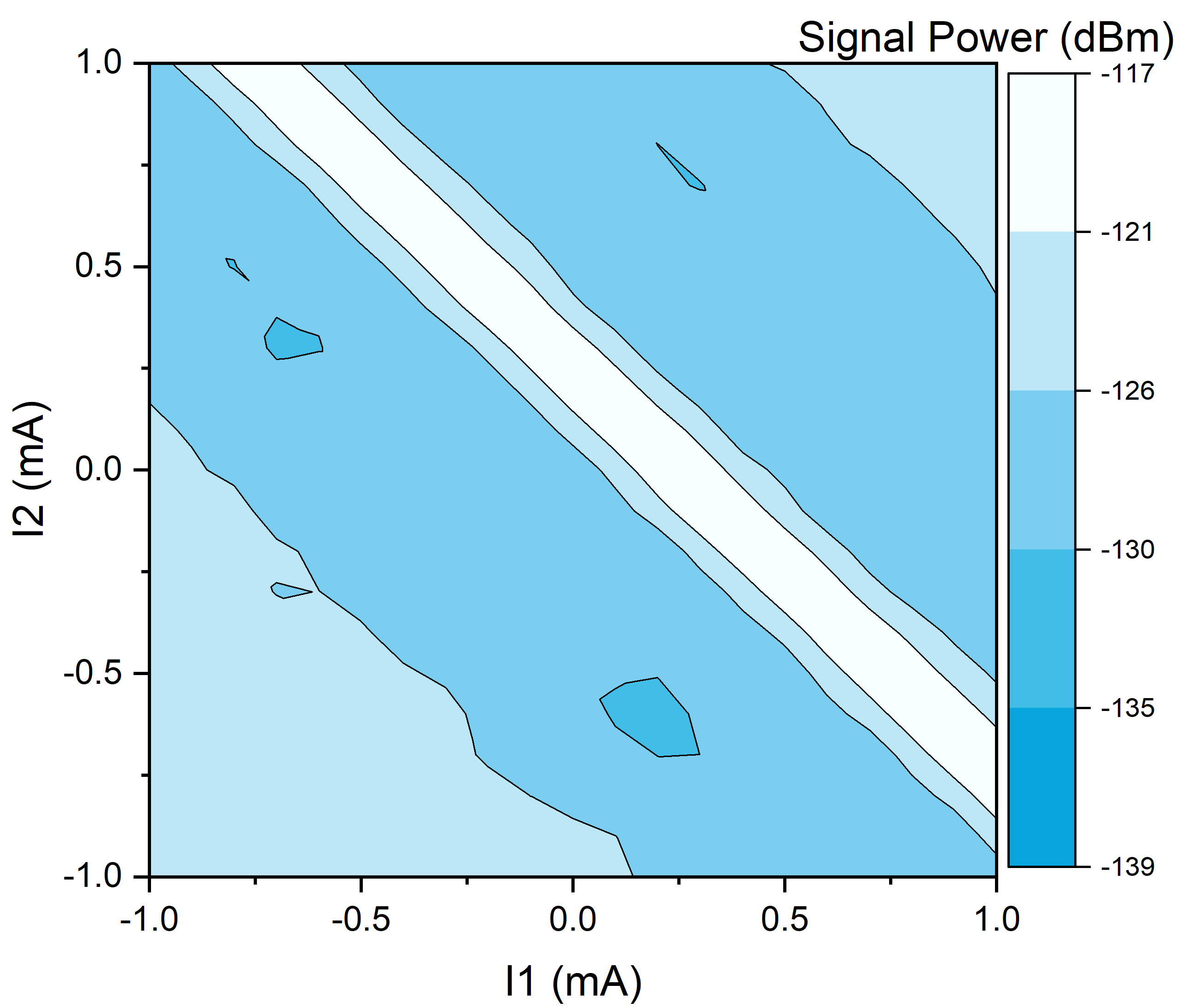}
 		\label{sweep_2_on}}
 	\caption{Comparison of signal power vs. bias currents with pump on and off. (a) On the segment corresponding to $\varphi_+\approx\pi/2$, with pump off, signal is mostly attenuated and converted into its second harmonic component. That's because, when $\varphi_{+}\approx\pi/2$, second order effect has maximum efficiency, giving a nice second harmonic generation. (b) With pump on, the segment corresponding to $\varphi_+\approx\pi/2$ is where signal is mostly amplified. In fact, in the on state, second order effect results into second order parametric process, giving a 3WM operation.}
 	\label{sweeps}
 \end{figure}

We compared current sweeps with pump ON and pump OFF, as in Fig. \ref{sweeps}. When pump is OFF, the bias points where second order effect is maximum correspond to maximum signal attenuation (see Fig. \ref{sweep_2_off}) because of second harmonic generation. Instead, with pump ON, the same bias points give maximum signal amplification (Fig. \ref{sweep_2_on}), corresponding to a 3WM process.
{At} these bias points, { and all over the 4GHz bandwidth in Fig. \ref{data}, the difference in signal amplitude between pump ON and OFF was at least 20dB, at pump OFF, the amplitude was always below measurements noise floor ($\sim$-139dBm). This mechanism could then be used as a broadband, non-resonant} \emph{Pump Driven on/off rf-Switch}. This could play an useful role in many applications, where very weak signals need to be accurately controlled.

Moreover, we verified the correct 3WM operation by checking the idler frequency: it was quite high in power, reaching values between -135dB and -123dB. STWPA was then generating a detectable idler signal, with a frequency given by 3WM energy conservation relation $f_i=f_p-f_s$ for each signal frequency value. Idler was detected both at higher and lower frequencies with respect to the signal ones. This means that STWPA could be potentially used also as { a coherent Up-Down converting rf-Mixer for weak signals, allowing the direct implementation of wave mixing at 4K or mK stage of a dilution refrigerator with independent control on mixing efficiency ($\beta$) and phase difference ($\gamma$) by means of external dc currents. We then believe that STWPA approach could help the transition to fully superconducting control electronics, replacing room temperature mixers.

These possible applications, in addition to amplifier one, make STWPA a very capable and versatile tool that could, in principle, perform a wide range of operation  within the same on-chip circuit}.
\section{Conclusions}
 We experimentally demonstrated the validity of our new symmetric approach to TWPA design using a novel Symmetric rf-SQUID. A 4GHz bandwidth at  3dB was measured at 4.2K operating temperature, with a peak value around 12.5dB. Thanks to the two-bias-degrees of freedom of the new STWPA, we were able to explore a wide operating range of the device, reaching gain up to 17dB. These results are promising substantial improvement over standard three-wave mixing TWPA, where a maximum 11 dB gain is achieved \cite{zorinexp}.
 The additional bias degree of freedom provided by the STWPA design allowed us to choose both $\beta$ and $\gamma$ around optimal values over a wide range of signal frequencies, also very close to the pump one. The STWPA also showed additional features, as high idler power and pump controlled switch. These suggest, respectively, that a STWPA could be used to make an { up/down-conversion mixer using the idler as output directly at the mK stage of a dilution cryostat, as well as a cold rf triggered broadband switch for few microwave photon signal control.}
 Moreover, the new symmetric rf-SQUID has value on its own. Preliminary results show that, with symmetric design, the rf-SQUID $\beta<1$ constraint can be bypassed. This could have many consequences in microwave Josephson circuits projecting. Its full dynamics are now being analyzed and will be experimentally tested.
 \section*{Acknowledgment}
 We want to thanks A.B. Zorin, M. Khabipov of PTB, I. V. Vernik, A.F. Kirichenko, D.Yohannes and all the Fabrication Team of HYPRES -SeeQC, Inc. for the help and suggestions they gave us during this work.  We want to thanks also R. Caruso, H.G. Ahmad, D. Massarotti, F. Tafuri, G. P. Pepe, D. Stornaiuolo, A. Andreone and A. Capozzoli for the invaluable help in starting these measurements in the new lab in Naples.
\bibliographystyle{IEEEtran}
\bibliography{./Symmetric_Traveling_Wave_Parametric_Amplifier_FINAL_VERSION_ARXIV}

\begin{thebibliography}{10}
\providecommand{\url}[1]{#1}
\csname url@samestyle\endcsname
\providecommand{\newblock}{\relax}
\providecommand{\bibinfo}[2]{#2}
\providecommand{\BIBentrySTDinterwordspacing}{\spaceskip=0pt\relax}
\providecommand{\BIBentryALTinterwordstretchfactor}{4}
\providecommand{\BIBentryALTinterwordspacing}{\spaceskip=\fontdimen2\font plus
\BIBentryALTinterwordstretchfactor\fontdimen3\font minus
  \fontdimen4\font\relax}
\providecommand{\BIBforeignlanguage}[2]{{%
\expandafter\ifx\csname l@#1\endcsname\relax
\typeout{** WARNING: IEEEtran.bst: No hyphenation pattern has been}%
\typeout{** loaded for the language `#1'. Using the pattern for}%
\typeout{** the default language instead.}%
\else
\language=\csname l@#1\endcsname
\fi
#2}}
\providecommand{\BIBdecl}{\relax}
\BIBdecl

\bibitem{jpc}
\BIBentryALTinterwordspacing
N.~Bergeal, R.~Vijay, V.~E. Manucharyan, I.~Siddiqi, R.~J. Schoelkopf, S.~M.
  Girvin, and M.~H. Devoret, ``Analog information processing at the quantum
  limit with a josephson ring modulator,'' \emph{Nature Physics}, vol.~6, pp.
  296 EP --, Feb 2010, article. [Online]. Available:
  \url{http://dx.doi.org/10.1038/nphys1516}
\BIBentrySTDinterwordspacing

\bibitem{flux_jpa}
\BIBentryALTinterwordspacing
T.~Yamamoto, K.~Inomata, M.~Watanabe, K.~Matsuba, T.~Miyazaki, W.~D. Oliver,
  Y.~Nakamura, and J.~S. Tsai, ``Flux-driven josephson parametric amplifier,''
  \emph{Applied Physics Letters}, vol.~93, no.~4, p. 042510, 2008. [Online].
  Available: \url{https://doi.org/10.1063/1.2964182}
\BIBentrySTDinterwordspacing

\bibitem{beyond_gb}
\BIBentryALTinterwordspacing
T.~Roy, S.~Kundu, M.~Chand, A.~M. Vadiraj, A.~Ranadive, N.~Nehra, M.~P.
  Patankar, J.~Aumentado, A.~A. Clerk, and R.~Vijay, ``Broadband parametric
  amplification with impedance engineering: Beyond the gain-bandwidth
  product,'' \emph{Applied Physics Letters}, vol. 107, no.~26, p. 262601, 2015.
  [Online]. Available: \url{https://doi.org/10.1063/1.4939148}
\BIBentrySTDinterwordspacing

\bibitem{mutus}
\BIBentryALTinterwordspacing
J.~Y. Mutus, T.~C. White, E.~Jeffrey, D.~Sank, R.~Barends, J.~Bochmann,
  Y.~Chen, Z.~Chen, B.~Chiaro, A.~Dunsworth, J.~Kelly, A.~Megrant, C.~Neill,
  P.~J.~J. O'Malley, P.~Roushan, A.~Vainsencher, J.~Wenner, I.~Siddiqi,
  R.~Vijay, A.~N. Cleland, and J.~M. Martinis, ``Design and characterization of
  a lumped element single-ended superconducting microwave parametric amplifier
  with on-chip flux bias line,'' \emph{Applied Physics Letters}, vol. 103,
  no.~12, p. 122602, 2013. [Online]. Available:
  \url{https://doi.org/10.1063/1.4821136}
\BIBentrySTDinterwordspacing

\bibitem{vissers}
\BIBentryALTinterwordspacing
B.~Ho~Eom, P.~K. Day, H.~G. LeDuc, and J.~Zmuidzinas, ``A wideband, low-noise
  superconducting amplifier with high dynamic range,'' \emph{Nature Physics},
  vol.~8, pp. 623 EP --, Jul 2012, article. [Online]. Available:
  \url{https://doi.org/10.1038/nphys2356}
\BIBentrySTDinterwordspacing

\bibitem{jos_meta}
\BIBentryALTinterwordspacing
M.~Trepanier, D.~Zhang, O.~Mukhanov, and S.~M. Anlage, ``Realization and
  modeling of metamaterials made of rf superconducting quantum-interference
  devices,'' \emph{Phys. Rev. X}, vol.~3, p. 041029, Dec 2013. [Online].
  Available: \url{https://link.aps.org/doi/10.1103/PhysRevX.3.041029}
\BIBentrySTDinterwordspacing

\bibitem{castellanos}
\BIBentryALTinterwordspacing
M.~A. Castellanos-Beltran and K.~W. Lehnert, ``Widely tunable parametric
  amplifier based on a superconducting quantum interference device array
  resonator,'' \emph{Applied Physics Letters}, vol.~91, no.~8, p. 083509, 2007.
  [Online]. Available: \url{https://doi.org/10.1063/1.2773988}
\BIBentrySTDinterwordspacing

\bibitem{yaakobi}
\BIBentryALTinterwordspacing
O.~Yaakobi, L.~Friedland, C.~Macklin, and I.~Siddiqi, ``Parametric
  amplification in josephson junction embedded transmission lines,''
  \emph{Phys. Rev. B}, vol.~87, p. 144301, Apr 2013. [Online]. Available:
  \url{https://link.aps.org/doi/10.1103/PhysRevB.87.144301}
\BIBentrySTDinterwordspacing

\bibitem{fract_twpa}
\BIBentryALTinterwordspacing
A.~A. Adamyan, S.~E. de~Graaf, S.~E. Kubatkin, and A.~V. Danilov,
  ``Superconducting microwave parametric amplifier based on a quasi-fractal
  slow propagation line,'' \emph{Journal of Applied Physics}, vol. 119, no.~8,
  p. 083901, 2016. [Online]. Available: \url{https://doi.org/10.1063/1.4942362}
\BIBentrySTDinterwordspacing

\bibitem{bell}
\BIBentryALTinterwordspacing
M.~T. Bell and A.~Samolov, ``Traveling-wave parametric amplifier based on a
  chain of coupled asymmetric squids,'' \emph{Phys. Rev. Applied}, vol.~4, p.
  024014, Aug 2015. [Online]. Available:
  \url{https://link.aps.org/doi/10.1103/PhysRevApplied.4.024014}
\BIBentrySTDinterwordspacing

\bibitem{zorinexp}
A.~B. Zorin, M.~Khabipov, J.~Dietel, and R.~Dolata, ``Traveling-wave parametric
  amplifier based on three-wave mixing in a josephson metamaterial,'' in
  \emph{2017 16th International Superconductive Electronics Conference (ISEC)},
  June 2017, pp. 1--3.

\bibitem{white}
\BIBentryALTinterwordspacing
T.~C. White, J.~Y. Mutus, I.-C. Hoi, R.~Barends, B.~Campbell, Y.~Chen, Z.~Chen,
  B.~Chiaro, A.~Dunsworth, E.~Jeffrey, J.~Kelly, A.~Megrant, C.~Neill, P.~J.~J.
  O'Malley, P.~Roushan, D.~Sank, A.~Vainsencher, J.~Wenner, S.~Chaudhuri,
  J.~Gao, and J.~M. Martinis, ``Traveling wave parametric amplifier with
  josephson junctions using minimal resonator phase matching,'' \emph{Applied
  Physics Letters}, vol. 106, no.~24, p. 242601, 2015. [Online]. Available:
  \url{https://doi.org/10.1063/1.4922348}
\BIBentrySTDinterwordspacing

\bibitem{macklin}
\BIBentryALTinterwordspacing
C.~Macklin, K.~O{\textquoteright}Brien, D.~Hover, M.~E. Schwartz,
  V.~Bolkhovsky, X.~Zhang, W.~D. Oliver, and I.~Siddiqi, ``A
  near{\textendash}quantum-limited josephson traveling-wave parametric
  amplifier,'' \emph{Science}, vol. 350, no. 6258, pp. 307--310, 2015.
  [Online]. Available: \url{http://science.sciencemag.org/content/350/6258/307}
\BIBentrySTDinterwordspacing

\bibitem{zorinth}
\BIBentryALTinterwordspacing
A.~B. Zorin, ``Josephson traveling-wave parametric amplifier with three-wave
  mixing,'' \emph{Phys. Rev. Applied}, vol.~6, p. 034006, Sep 2016. [Online].
  Available: \url{https://link.aps.org/doi/10.1103/PhysRevApplied.6.034006}
\BIBentrySTDinterwordspacing

\bibitem{barone_squid}
\BIBentryALTinterwordspacing
A.~Barone and G.~Paternò, \emph{Physics and Applications of the Josephson
  Effect}.\hskip 1em plus 0.5em minus 0.4em\relax Wiley-Blackwell, 2005,
  ch.~13, pp. 383--445. [Online]. Available:
  \url{https://onlinelibrary.wiley.com/doi/abs/10.1002/352760278X.ch13}
\BIBentrySTDinterwordspacing

\end{thebibliography}
\end{document}